\documentclass[aps,prd,twocolumn,groupedaddress]{revtex4-1}
\usepackage{graphicx}
\include{amssym}

\begin{document}
\title{Gauge covariant approach to the electroweak interactions of mesons in the Nambu-Jona-Lasinio model with spin-1 states}

\author{A. A. Osipov$^{1,2}$}
\email[]{aaosipov@jinr.ru}

\author{B. Hiller$^{3}$}
\email[]{brigitte@fis.uc.pt}

\author{P. M. Zhang$^{1,4}$}
\email[]{zhpm@impcas.ac.cn}

\affiliation{$^1$Institute of Modern Physics, Chinese Academy of Sciences, Lanzhou 730000, China}
\affiliation{$^2$Bogoliubov Laboratory of Theoretical Physics, Joint Institute for Nuclear Research, Dubna, 141980, Russia}
\affiliation{$^3$CFisUC, Department of Physics, University of Coimbra, P-3004-516 Coimbra, Portugal}
\affiliation{$^4$School of Physics and Astronomy, Sun Yat-sen University, Zhuhai 519082, China}

\begin{abstract}
An extended Nambu-Jona-Lasinio (NJL) model with chiral group $U(2)\times U(2)$ and spin-0 and spin-1 four quark interactions is used to develop the gauge covariant approach to the diagonalization of the $\pi-a_1$ mixing in the presence of electroweak forces. This allows for manifestly gauge covariant description of both the non-anomalous and anomalous parts of the effective Lagrangian. It is shown that in the non-anomalous sector the theory is equivalent to the standard non-covariant approach. 
\end{abstract}

\maketitle

\section{Introduction}
In the Nambu - Jona-Lasinio model \cite{Nambu61a,Nambu61b} the mixing of pseudoscalar and axial-vector bound states, $p -a_\mu$ mixing \cite{Osipov85,Morais17}, is an inevitable concomitant of spontaneous chiral symmetry breaking. Similar mixing takes also place in the majority of chiral effective Lagrangians with spin-1 states \cite{Gasiorovicz69,Meissner88,Bando88,Birse96} [the alternative option is the description of massive spin-1 fields in terms of antisymmetric tensors \cite{Gasser84,Schaden87,Ecker89plb,Ecker89npb}; this approach is free from $p-a_\mu$ mixing, but is less popular in phenomenological particle physics]. 

Our main concern here are theories with the usual vector field formulation in which chirality is treated as an ordinary linear symmetry broken by the vacuum. In particular, the NJL model is used to elucidate the approach. Our main goal is a covariant $p-a_\mu$ diagonalization of the effective meson Lagrangian in presence of $SU(2)_L\times U(1)_R$ gauge invariant electroweak interactions of quarks, and as a consequence, of mesons. 

Since the gauge group is a subgroup of the chiral $SU(2)_L\times SU(2)_R$ group, the latter is chosen as a symmetry of the NJL quark Lagrangian. This simplified approach [there are no strange quarks] makes calculations more transparent. According to this plan, we will focus mainly on the most important points: the clear formulation of the main idea, the gauge transformation laws of the fields involved, the structure of the non-anomalous part of the total effective action and the proof of its equivalence to the standard one \cite{Ebert82,Ebert83,Volkov84,Volkov86,Ebert86}. The material presented here will be considered in detail in the forthcoming paper \cite{Osipov18arx}.             

The reason why a covariant $p-a_\mu$ diagonalization is preferable to the standard (non-covariant) one 
\begin{equation}
\label{noncd}
a_{\mu}\to a_{\mu}+ \kappa m \partial_\mu p
\end{equation}
is easy. In presence of gauge symmetry the ordinary derivative $\partial_\mu p$ violates it. In (\ref{noncd}) $\kappa$ is a constant, $m$ is the constituent quark mass, $a_\mu =a_{\mu a}\tau_a, \,\, p=p_a\tau_a $ with $a=0,1,2,3$,  $\tau_0=1$, and $\tau_i$ $(i=1,2,3)$ are the Pauli matrices. Recently this has been demonstrated for the electromagnetic interactions \cite{Osipov18jl,Osipov18vd}. In particular,  it has been shown that the amplitude of the anomalous radiative $a_1(1260)\to\gamma\pi^+\pi^-$ decay is not gauge invariant if evaluated in the standard approach. 

One may ask if these violations of $U(1)$ gauge symmetry may only happen in the anomalous sector of the theory. The answer is positive, and in this paper we explain the reason. Additionally, we show that there aren't observable physical consequences between covariant and non-covariant approaches when the non-anomalous electroweak interactions of mesons are considered. It means that the standard approach, which relates to the phenomenologically very successful vector/axial-vector dominance picture, first introduced by \cite{Nambu57,Frazer60,Gell-Mann61,Gell-Mann62,Nambu62,Kroll67}, remains unaltered for non-anomalous processes. 

Unfortunately, we cannot extend this picture to the description of anomalous processes. The $a_1(1260)\to \gamma \pi^+\pi^-$ decay is probably not the only case where the vector meson dominance (VMD) approach fails [another example has been found in \cite{Kugo85}, where the authors came to the conclusion that the "complete vector meson dominance" hypotesis of photon couplings is invalid in either $\pi^0\to\gamma \gamma$ or $\gamma\to\pi\pi\pi$ processes]. As obtained in \cite{Osipov18jl,Osipov18vd}, the reason is related to gauge symmetry which is violated due to missing diagrams in the cases that the $p-a_\mu$ mixing is not diagonalized in a covariant way. Indeed, in the $a_1\to\gamma\pi^+\pi^-$ decay, only anomalous vertices are responsible for emission of the single photon: $a_1\rho^0\gamma,\ \rho^\pm \pi^\mp\gamma$ and the quark box diagram $a_1\gamma\pi^+\pi^-$. In both the standard and the covariant approaches these vertices are of the VMD type. As a result, both approaches give the $U(1)$ gauge symmetry breaking result for the amplitude. However, the covariant approach, due to nonlinear interactions induced through the covariant version of replacement (\ref{noncd}) [see Eqs. (\ref{da}-\ref{ymu}) below], possesses the additional anomalous non-VMD triangle diagram with emission of the photon together with the pion in one of the vertices. This protects the gauge symmetry of the amplitude but deviates from the VMD picture. Therefore, there is a definite benefit from the covariant formulation of the $p-a_\mu$ diagonalization which we extend here to the entire electroweak sector.

\section{Main idea}

Let us discuss now the main idea of our approach. Instead of (\ref{noncd}), we want to construct a manifestly gauge and chiral-covariant replacement. Such replacement should not ruin the standard transformation laws of fields. Under standard we mean fundamental transformations of quark fields and adjoint transformations of meson fields. We use a model with linear realization of chiral symmetry and therefore should consider the meson states to be chiral partners. As a consequence of this fact, if the axial field changes as $a_\mu\to a_\mu+\Delta a_\mu$ the corresponding replacement in the vector field $v_\mu\to v_\mu +\Delta v_\mu$ is such that the term $\Delta v_\mu$ is a chiral partner of $\Delta a_\mu$. Fortunately, for the chiral group this step is known \cite{Osipov00,Osipov02}. In the spontaneously broken phase, the following modification of spin-1 fields is required
\begin{eqnarray}
\label{nongcov}
a_\mu&\to& a_\mu +\frac{\kappa}{2} \left(\{p,\partial_\mu \bar s\} -\{\bar s,\partial_\mu p \} \right),  \\
\label{nongcovv}
v_\mu&\to& v_\mu +i\,\frac{\kappa}{2}\left([p,\partial_\mu p] + [\bar s, \partial_\mu \bar s] \right),
\end{eqnarray}
where the scalar field $s$ is given by $\bar s =s-m$. 

Notice that the two bilinear combinations of scalar and pseudoscalar fields transform like axial-vector and vector fields and are chiral partners with respect to the linear transformations of the chiral group. One can also see that the linear part of Eq. (\ref{nongcov}) is identical to the replacement (\ref{noncd}). So, the price for the apparent chiral-covariant form of the $p-a_\mu$ diagonalization are the new terms bi-linear in the fields which will not affect the $S$-matrix elements of the theory on mass shell due to the Chisholm-Kamefuchi-O'Raifeartaigh-Salam theorem \cite{Chisholm61,Salam61}. Nonetheless, this representation is more suitable for our second step. 

Our second step is motivated by the observation that the above replacements will not be gauge covariant, if the theory possesses also the local gauge symmetry. In this case, one should replace in (\ref{nongcov}) and (\ref{nongcovv}) the usual derivatives of the pseudoscalar and scalar fields by the $SU(2)_L\times U(1)_R$ gauge covariant ones 
\begin{eqnarray}
\label{dp}
{\cal D}_\mu p &=&  \partial_\mu p-i [N_\mu, p]-\{K_\mu, \bar s \}, \\
\label{ds}
{\cal D}_\mu \bar s &=& \partial_\mu \bar s   -i [N_\mu, \bar s]+\{K_\mu, p \}.
\end{eqnarray}
The auxiliary fields $N_\mu$ and $K_\mu$ are defined as follows
\begin{eqnarray}
&& N_\mu=\frac{1}{2}\left(gA_\mu+g'B_\mu T_3\right), \\
&&K_\mu =\frac{1}{2}\left(gA_\mu -g'B_\mu T_3\right),
\end{eqnarray}
where $A_\mu = A_iT_i$ and $B_\mu$ are gauge fields of the $SU(2)_L$ and $U(1)_R$ groups of local transformations, correspondingly; $T_i = \tau_i/2$, and the couplings of electroweak interactions $g$ and $g'$ are related to the Weinberg angle $\theta_W$ as follows: $\cos \theta_W = g/\sqrt{g^2+(g')^2}$, and $\sin \theta_W = g'/\sqrt{g^2+(g')^2}$. The physical variables $Z_\mu, {\cal A}_\mu$ and $W_\mu^\pm$  of the gauge fields can be introduced through the following transformations
\begin{eqnarray}
&& Z_\mu =  \cos \theta_W A_\mu^3 -\sin\theta_W B_\mu ,  \nonumber  \\
&& {\cal A}_\mu =  \sin\theta_W A_\mu^3 + \cos \theta_W B_\mu , \nonumber \\
&&W_\mu^\pm=\left(A_\mu^1\mp iA_\mu^2\right)/\sqrt{2}.
\end{eqnarray}

Both derivatives (\ref{dp}) and (\ref{ds}) belong to the adjoint representation of the gauge group and are gauge covariant, because they transform like the ordinary pseudoscalar and scalar fields under $SU(2)_L\times U(1)_R$. 

The replacements (\ref{nongcov}) and (\ref{nongcovv}) are rendered chiral and gauge covariant through the modifications 
\begin{equation}
\label{da}
a_\mu\to a_\mu +\frac{\kappa}{2}\,Y_\mu , \quad  v_\mu\to v_\mu +\frac{\kappa}{2}\,X_\mu,
\end{equation}
where $X_\mu$ and $Y_\mu$ must have the form
\begin{eqnarray}
\label{xmu}
X_\mu &=& i\left(   [p,{\cal D}_\mu p]+[\bar s, {\cal D}_\mu\bar s] \right), \\
\label{ymu}
Y_\mu &=& \{p, {\cal D}_\mu\bar s \}-\{\bar s, {\cal D}_\mu p \},
\end{eqnarray}
to be chiral and gauge covariant objects. One can also check that $X_\mu^\dagger =X_\mu$ and $Y_\mu^\dagger =Y_\mu$. 

The covariant replacement (\ref{da}) is the core element of our approach. 

\section{Realization}
\label{L}
To understand how this works, let us apply the above idea to the theory described by the extended $U(2)\times U(2)$ NJL Lagrangian in presence of electroweak interactions. 
 
For that we need the vacuum-to-vacuum amplitude of the model. We will write it in the Nambu-Goldstone phase, where the quark fields $q(x)$ get their constituent masses $m=\mbox{diag}(m_u,m_d)$ as a result of spontaneous chiral symmetry breaking. In the non-symmetric vacuum, the physical spectrum is represented by $q\bar q$ bound states: $s_a\propto\bar q\tau_a q$, $p_a\propto\bar q i\gamma_5\tau_a q$, $v_{\mu a}\propto\bar q \gamma_\mu \tau_a q$, and $a_{\mu a}\propto\bar q\gamma_\mu\gamma_5\tau_a q$. Hence, it is convenient to introduce meson variables in the corresponding functional integral explicitly. This can be done by applying a Hubbard-Stratonovich transformation to rewrite the nonlinear four-quark interactions in terms of Yukawa interactions of quarks with auxiliary boson fields. The corresponding vacuum-to-vacuum amplitude is then given by
\begin{eqnarray}
\label{smat3}
&&S [{\cal A}_\mu, Z_\mu, W_\mu^\pm ]=\int [dq] [d\bar q] [ds_a] [dp_a] [dv_{a\mu} ]
   [d{a}_{a\mu} ]  \nonumber \\
&&\times\exp i\!\!\int\!\! d^4x\left(\bar q\, D_mq +\mathcal L_{M0} +\mathcal L_{M1} +{\cal L}_{EW}\right),
\end{eqnarray}
where $D_m$ is the Dirac operator in presence of background fields
\begin{equation}
\label{Dm}
D_m = i\gamma^\mu\mathcal D_\mu -m+s+i\gamma_5 p +\gamma^\mu v_\mu
+ \gamma^\mu\gamma_5 a_\mu ,
\end{equation}
where $s=s_a\tau_a,\, p=p_a\tau_a,\, v_{\mu } =v_{a\mu }\tau_a,\, a_\mu =a_{a\mu }\tau_a$ are the scalar, pseudoscalar, vector and axial vector fields correspondingly. 

The gauge covariant quark derivative has the form
\begin{equation}
{\cal D}_\mu q = \left[\partial_\mu -igA_\mu P_L -ig'B_\mu (Q-T_3P_L) \right]q,
\end{equation}
where the matrix $Q=T_3+Y_L=1/2(\tau_3 +1/3)$ describes the electromagnetic charges of $u$ and $d$ quarks in relative units of the proton charge $e>0$; $P_{L,R}=(1\mp\gamma_5)/2$.

$\mathcal L_{M0}$ and $\mathcal L_{M1}$ describe the mass part of the spin-0 and spin-1 fields.
\begin{eqnarray}
\label{Mpart0}
\mathcal L_{M0} &=&-\frac{1}{4G_S}\mbox{tr} \left[ (s -m+\hat m)^2+p^{2}\right],     \\
\label{Mpart1}
\mathcal L_{M1}&=&\frac{1}{4G_V} \mbox{tr} \left( v_\mu^{2}+a_\mu^{2}\right),
\end{eqnarray}
where $G_S$ and $G_V$ are the couplings of the spin-less and spin-1 four-quark interactions; the matrix $\hat m =\mbox{diag}(\hat m_u, \hat m_d)$ collects the masses of current quarks, and the trace is taken over flavor indices. 

The Lagrangian density ${\cal L}_{EW}$ contains only the electroweak part of the theory. In the following, we will not use it in our calculations. For this reason the corresponding expression is not given.

The evaluation of the path integral (\ref{smat3}) over quark fields leads to $p-a_\mu$ mixing through the one quark loop amplitude. The mixing term occurs to be proportional to $\mbox{tr}(a_\mu\partial_\mu p)$. To avoid this unwanted contribution we make the replacement of variables (\ref{da}) in the path integral. After integration over quarks, the coupling $\kappa$ is unambiguously fixed by the diagonalization condition. The procedure (\ref{da}) has the following consequences. 

First, it changes the form of the Dirac operator by introducing a set of linear and nonlinear interactions of mesons and gauge fields with quarks
\begin{eqnarray}
\label{Dm}
D_m\to D_m&=&i\gamma^\mu{\cal D}_\mu +\bar s+i\gamma_5p +\gamma^\mu (v_\mu +\gamma_5 a_\mu) \nonumber \\
&+&\frac{\kappa}{2}\gamma^\mu (X_\mu + \gamma_5 Y_\mu).
\end{eqnarray}
Notice that after such replacement $D_m$ does not lose its covariant transformation properties under the action of chiral and gauge groups. This saves the theory from all sorts of possible symmetry breaking effects. 

Eq. (\ref{Dm}) can be rewritten in a more convenient way as follows
\begin{equation}
D_m=i\gamma^\mu\partial_\mu +\bar s +i\gamma_5 p +\gamma^\mu\left(\Gamma_\mu^{(V)}+\gamma_5\Gamma_\mu^{(A)}\right),
\end{equation}
where
\begin{eqnarray}
\Gamma_\mu^{(V)}&=&v_\mu+eQ{\cal A}_\mu +\frac{gZ_\mu}{2\cos\theta_W}\left(T_3-Q\sin^2\theta_W\right) \nonumber \\
&+& \frac{g}{2}\left(T_+W_\mu^++T_-W_\mu^-\right) +\frac{\kappa}{2}\,X_\mu ,\\
\Gamma_\mu^{(A)}&\!\!\!=\!\!\!&a_\mu-\frac{gT_3Z_\mu}{2\cos\theta_W} -\frac{g}{2}\left(T_+W_\mu^++T_-W_\mu^-\right) \nonumber \\
&+& \frac{\kappa}{2}\,Y_\mu ,
\end{eqnarray}
$T_\pm = (T_1\pm iT_2)/\sqrt{2}$, and $\sin^2\theta_W=0.23$.

Second, after replacement (\ref{da}), the Lagrangian density (\ref{Mpart1}) gets some vertices with interactions. We will consider them below, but first let us eliminate the solitary interactions of the photon, $Z$ and $W^\pm$ bosons with quarks. This is in line with the noted idea of vector/axial-vector meson dominance. We suggest for that a very simple procedure which is markedly different from the quite complicated method used in \cite{Ebert82,Ebert83,Volkov84,Volkov86,Ebert86}. 

Indeed, the following simple replacements yield the desired result
\begin{eqnarray}
\label{vdr}
v_\mu&\to& v_\mu -eQ{\cal A}_\mu -\frac{gZ_\mu}{2\cos\theta_W}\left(T_3-Q\sin^2\theta_W\right) \nonumber\\
&-&\frac{g}{2} \left(T_+W_\mu^++T_-W_\mu^-\right) \\
\label{avd}
a_\mu&\to& a_\mu +g_A\, \frac{g}{2}\left(\frac{T_3Z_\mu}{\cos\theta_W} +T_+W_\mu^++T_-W_\mu^-\right),
\end{eqnarray}
where $g_A=1-2\kappa m^2$. 

These replacements change $\Gamma_\mu^{(V,A)} $ accordingly
\begin{eqnarray}
\label{gammav}
\Gamma_\mu^{(V)}&\to&\Gamma_\mu^{(V)}=v_\mu +\frac{\kappa}{2}\,X_\mu ,\\
\label{gammaa}
\Gamma_\mu^{(A)}&\to&\Gamma_\mu^{(A)}=a_\mu +\frac{\kappa}{2}\,Y_\mu \nonumber \\
     &-& \kappa gm^2\left(\frac{T_3Z_\mu}{\cos\theta_W}+T_+W_\mu^++T_-W_\mu^-\right).
\end{eqnarray}
One can see that the new $\Gamma_\mu^{(V,A)}$ functions still contain the gauge fields inside the covariant derivatives in $X_\mu$ and $Y_\mu$. This is a signal that the theory may deviate from the vector and axial-vector dominance. Notice, that such deviation, if it happens, is a direct result of the covariant replacement (\ref{da}).

Now we return to the Lagrangian density $\mathcal L_{M1}$. With redefinitions (\ref{da}), (\ref{vdr}) and (\ref{avd}), which we summarize as $v_\mu\to\tilde v_\mu$, and $a_\mu\to\tilde a_\mu$, the original mass term (\ref{Mpart1}) becomes
\begin{eqnarray}
\label{vt2}
\mbox{tr}&&\left(\tilde v_\mu^2\right) = \mbox{tr}\left[\left(v_\mu + \frac{\kappa}{2}\, \xi_\mu \right)  \left(v_\mu + \frac{\kappa}{2} \,\xi_\mu +\kappa\,\Xi_\mu \right) \right] \nonumber \\
&& -2e{\cal A}_\mu \left(\frac{v_{\mu 0}}{3}+v_{\mu 3} +  \frac{\kappa}{2} \,\xi_{\mu 3} \right)\nonumber\\
&&+\frac{gZ_\mu}{\cos\theta_W}\left[ \frac{v_{\mu 0}}{3}\sin^2\theta_W - \left( v_{\mu 3} + \frac{\kappa}{2}\,\xi_{\mu 3}\right)\cos^2\theta_W  \right] \nonumber \\
&& -g\left[ W_\mu^+  \left( v_{\mu}^- + \frac{\kappa}{2}\,\xi_{\mu -}\right) +
                 W_\mu^-  \left( v_{\mu}^+ + \frac{\kappa}{2}\,\xi_{\mu +}\right) \right] \nonumber \\
&& +\left( \kappa\, \Xi_{\mu +} - \frac{g}{2}\,W_\mu^+ \right)\left( \kappa\, \Xi_{\mu -} - \frac{g}{2}\,W_\mu^- \right) \nonumber\\
&&+\frac{1}{18}\left( \frac{gZ_\mu}{2\cos\theta_W}\sin^2\theta_W - e {\cal A}_\mu  \right)^2 \nonumber \\
&&+\frac{1}{2} \left( \kappa\,\Xi_3 - e {\cal A}_\mu -\frac{g}{2} \, Z_\mu \cos\theta_W  \right)^2 ,
\end{eqnarray}
and
\begin{eqnarray}
\label{at2}
\mbox{tr}&&\left(\tilde a_\mu^2\right) = \mbox{tr}\left(a_\mu +\frac{\kappa}{2}\, Y_\mu\right)^2 +
\frac{gg_A Z_\mu}{\cos\theta_W} \left( a_{\mu  3} +\frac{\kappa}{2}\, Y_{\mu 3}  \right)    \nonumber\\
&&+gg_A \left[ W_\mu^+  \left( a_{\mu}^- + \frac{\kappa}{2}\, Y_{\mu -}\right) +
                 W_\mu^-  \left( a_{\mu}^+ + \frac{\kappa}{2}\, Y_{\mu +}\right) \right] \nonumber \\
&&+\frac{1}{4}\,g^2g_A^2 \left(W_\mu^+ W_\mu^- +\frac{Z_\mu^2}{2\cos^2\theta_W}   \right).
\end{eqnarray}
Here the following conventions [additionally to Eqs. (\ref{xmu}) and (\ref{ymu})] were adopted:
\begin{eqnarray}
X_\mu &=& \xi_\mu + \Xi_\mu , \quad \xi_\mu =i\left([p,\partial_\mu p]+[s, \partial_\mu s]\right), \nonumber\\
Y_\mu &=&\zeta_\mu + \Psi_\mu, \quad \zeta_\mu= \{p, \partial_\mu \bar s \}-\{\bar s, \partial_\mu  p \}.
\end{eqnarray}
All 4-vectors are elements of the Lie algebra of the $U(2)$ group, e.g., $\xi_\mu=\xi_{\mu a}\tau_a$. 
Extracting from (\ref{vt2})  and (\ref{at2}) the terms of the second order in powers of fields, and taking into account  the standard redefinition of the meson fields, see Eqs. (\ref{Rsp})-(\ref{Rvach}), we arrive at the conventional vector and axial-vector meson dominance results
\begin{eqnarray}
\label{vmd}
{\cal L}_\gamma &=&-e\,\frac{m_\rho^2}{g_\rho}{\cal A}_\mu\left(\frac{\omega_\mu}{3}+\rho^0_\mu \right), \\
\label{wd}
{\cal L}_{W^\pm} &=&\frac{g}{2}\,W_\mu^\pm\left[\frac{m_\rho^2}{g_\rho} \left(a_{1\mu}^\mp -\rho_\mu^\mp \right)+f_\pi\partial_\mu \pi^\mp \right],
\end{eqnarray}
including mixing of the $Z$ boson with the neutral vector, the axial-vector mesons and the pion
\begin{eqnarray}
\label{zd}
{\cal L}_Z&=&\frac{gZ_\mu}{2\cos\theta_W}\left[\frac{m_\rho^2}{g_\rho} \left(\frac{\omega_\mu}{3}\sin^2\theta_W-\rho^0_\mu \cos^2\theta_W +a_{1\mu}^0 \right) \right. \nonumber\\
&+&\left. f_\pi \partial_\mu\pi^0 \right].
\end{eqnarray}
Eq. (\ref{vmd}) summarizes the $U(2)$ field-current identities, describing electromagnetic VMD phenomena. Eq. (\ref{wd}) represents the corresponding result for the charged hadronic weak current. Eq. (\ref{zd}) shows weak mixing of neutral spin-1 states with the $Z$ boson. 

Let us integrate now in (\ref{smat3}) over the quark fields. The corresponding Gaussian path integral accounts for the one-quark-loop contribution to the effective action. The result is given by the non-local functional determinant. In particular, the contribution of this chiral determinant to the non-anomalous part of the effective action is given by
\begin{equation}
\label{seff}
         S= - \frac{i}{2}\,\mbox{Tr}\ln D_m^\dagger D_m =i\mathcal L,
\end{equation}
where the trace "Tr" should be calculated over color, Dirac, flavor indices and it also includes the integration over coordinates of the Minkowski space-time. 

The consistent approximation scheme to obtain from the non-local chiral determinant the long wavelength (low-energy) expansion for the effective quasilocal action $S$ is the Schwinger-DeWitt technique \cite{Schwinger54,DeWitt65,Ball89}. In the following, we will restrict ourselves to the first and second-order Seeley-DeWitt coefficients in such expansion. These coefficients accumulate the divergent part of the effective action, which is regularized here by a covariant ultraviolet cutoff $\Lambda$. Let us recall that the result of such calculations is well known [in the sense that the only difference between the expression for $D_m$ obtained in \cite{Osipov17aph} and $D_m$ here is the replacement of the vector $v_\mu$ and axial-vector field $a_\mu$ by $\Gamma_\mu^{(V)}$ and $\Gamma_\mu^{(A)}$ from Eqs. (\ref{gammav}) and (\ref{gammaa}) correspondingly]. Thus, we can use that result by writing
\begin{eqnarray}
\label{LSD1}
{\cal L}\to &{\cal L}_{SD}&=I_2\,\mbox{tr}\left\{ (\bigtriangledown_\mu s)^2 +  (\bigtriangledown_\mu p )^2   \right. \nonumber \\
&-&\left.\!\!\! (s^2-2m s+p^2)^2 - \frac{1}{3}(v_{\mu\nu} ^2 +a_{\mu\nu}^2)\right\}.
\end{eqnarray}
The overall factor $I_2=N_c f(\Lambda^2/m^2)$, with $N_c$ being the number of color degrees of freedom; the function $f$ is defined by the regularization. The other notations are 
\begin{eqnarray}
v_{\mu\nu}&=& \partial_\mu \Gamma_\nu^{(V)}-\partial_\nu \Gamma_\mu^{(V)}-i[\Gamma_\mu^{(V)},\Gamma_\nu^{(V)}]-i[\Gamma_\mu^{(A)},\Gamma_\nu^{(A)}], \nonumber\\
a_{\mu\nu}&=& \partial_\mu \Gamma_\nu^{(A)}-\partial_\nu \Gamma_\mu^{(A)}-i[\Gamma_\mu^{(V)},\Gamma_\nu^{(A)}]+i[\Gamma_\nu^{(V)},\Gamma_\mu^{(A)}], \nonumber \\
\bigtriangledown_\mu s&=&\partial_\mu s -i[\Gamma_\mu^{(V)}, s]-\{ \Gamma_\mu^{(A)} ,p\}, \nonumber \\
\bigtriangledown_\mu p&=& \partial_\mu p -i [\Gamma_\mu^{(V)} ,p]+\{\Gamma_\mu^{(A)},\bar s \}.
\end{eqnarray}

The total effective Lagrangian density of the meson and gauge fields is obtained by adding to the Schwinger-DeWitt part ${\cal L}_{SD}$ (\ref{LSD1}) the mass term of the spin-0 fields $\mathcal L_{M0}$ (\ref{Mpart0}), and the contribution arising from the mass term of spin-1 fields, after all changes of vector fields are taken into account [the latter correspond to the substitution $\mathcal L_{M1} \to \mathcal L_1$] and inclusion of the electroweak term ${\cal L}_{EW}$ 
\begin{equation}
\label{Ltot}
{\cal L}_{tot}={\cal L}_{SD} + {\cal L}_0+ {\cal L}_1 +{\cal L}_{EW},
\end{equation}
where 
\begin{equation}
{\cal L}_0 = - \frac{\hat m  \mbox{tr}   \left(s^2+p^2\right)}{4mG_S}  , \ \   {\cal L}_1 = \frac{\mbox{tr}\left(\tilde v_\mu^2+\tilde a_\mu^2\right)}{4G_V}.   
\end{equation}

Some comments about formula (\ref{Ltot}) are still in order. To get this Lagrangian density $\mathcal L_0$ we have used the gap equation $m-\hat m= mG_S I_1$, where $I_1=N_c f_1(m^2,\Lambda^2/m^2)$ is defined by the regularization. This equation is the condition that removes the unwanted $s_0$-tadpole contribution. 

The Lagrangian density ${\cal L}_{tot}$ does not contain $p-a_\mu$ mixing. This is because of the cancellation which occurs between two different contributions to the non diagonal $p-a_\mu$ mixing term in ${\cal L}_{tot}$. It restricts the numerical value of the parameter $\kappa$ to be
\begin{equation}
\label{pac}
\frac{1}{2\kappa} = m^2 +\frac{1}{16G_V I_2}.
\end{equation}

The free part of the Lagrangian density ${\mathcal L}_{tot}$ must display the canonical form. This can be achieved by the redefinition of the field variables
\begin{eqnarray}
\label{Rsp}
&& s =g_\sigma \sigma, \qquad  \ \   p=g_\pi \pi, \\
\label{Rvan}
&& v^0_\mu =\frac{g_\rho}{2}\omega_\mu, \quad  a^0_\mu =\frac{g_\rho}{2} f_{1\mu}, \\
\label{Rvach}
&&\vec v_\mu=\frac{g_\rho}{2}\vec\rho_\mu, \quad \vec a_\mu =\frac{g_\rho}{2}\vec a_{1\mu}.
\end{eqnarray}
The effective constants $g_\sigma , g_\pi , g_\rho$ and masses of meson states are functions of the $I_2$ and the constant $Z=g_A^{-1}$
\begin{eqnarray}
\label{couplings}
&&g_\sigma^2=\frac{1}{4I_2},\quad  g_\pi^2=Zg_\sigma^2,\quad  g_\rho^2=6g_\sigma^2, \\
&&m_\pi^2=\frac{\hat mg_\pi^2}{mG_S},\quad m_\sigma^2=4m^2+g_A m_\pi^2, \\
\label{mrho}
&&m_\rho^2=m_\omega^2=\frac{3}{8G_VI_2}, \\
\label{mrel}
&&m_{a_1}^2=m_{f_1}^2=m_\rho^2+6m^2.
\end{eqnarray}

\section{Hidden vector and axial-vector dominance}

The final topic to be considered here is whether the vector/axial-vector meson dominance is preserved in the covariant approach. The Lagrangian (\ref{Ltot}) has obviously vertices with direct emission of the gauge bosons by the quark-antiquark pair. Thus, there are two alternatives: either a gauge covariant description hides the dominance picture or the latter is not supported by the covariant approach. The case of the anomalous $a_1\to\gamma\pi\pi$ decay is a clear signal in favour of the second alternative. But what can one say about the non-anomalous processes?            
 
 To answer this question we will use gauge symmetry arguments. The $SU(2)_L\times U(1)_R$ gauge group acts on the quark fields by the following infinitesimal transformations
\begin{equation}
\delta q_L = i(\omega +e\alpha Y_L) q_L,   \quad \delta q_R = ie\alpha Qq_R,  
\end{equation}
where $q_R=P_R q$, $q_L=P_L q$, and  $\alpha$ and $\omega =\omega_iT_i$ are the local parameters of the $U(1)_R$ and $SU(2)_L$ gauge transformations. Now, using the quark structure of bound states, we find that meson fields transform as follows 
\begin{eqnarray}
\label{gts}
&&\delta \bar s = i [\theta , \bar s]+\{ \beta , p \} , \\
&& \delta p = i [\theta , p]-\{\beta , \bar s \}, \\
\label{v}
&&\delta  v_\mu = i [\theta , v_\mu ]+ i [\beta , a_\mu ], \\
\label{a}
&&\delta a_\mu = i [\theta , a_\mu ]+i [\beta , v_\mu ].
\end{eqnarray}
where the following new set of local parameters has been introduced
\begin{equation}
\theta =\frac{1}{2}\left(\omega + e\alpha T_3 \right), \quad
\beta =-\frac{1}{2}\left(\omega -e\alpha T_3  \right).
\end{equation}
In these notations the gauge transformations remind us of the usual chiral laws, but with local parameters. The gauge covariant diagonalization does not change these properties. As a result, after diagonalization (\ref{da}) the theory remains gauge invariant by construction.  
 
In the standard approach, the meson fields before diagonalization (\ref{noncd}) transform covariantly, i.e., in accord with Eqs. (\ref{gts})-(\ref{a}). However, the non-covariant diagonalization changes the laws for vector and axial-vector fields, keeping the transformations of spin-0 fields unchanged. So we obtain from (\ref{noncd})   
\begin{eqnarray}
\label{gtra}
\delta a_\mu &=& i[\theta, a_\mu ]+ i [\beta, v_\mu] -i\kappa m [\partial_\mu\theta , p] \nonumber \\
&+&\kappa m \partial_\mu \{\beta, \bar s\}, \\
\label{gtrv}
\delta v_\mu &=& i[\theta, v_\mu ]+ i [\beta, a_\mu] +i\kappa m [\beta , \partial_\mu p]. 
\end{eqnarray}  
These transformations still belong to the $SU(2)_L\times U(1)_R$ gauge group. Indeed, the sequential infinitesimal transformations $1, 2 , 1^{-1}, 2^{-1}$ corresponding to the commutator $\delta_{[1 2]}= [\delta_1, \delta_2]$ now acquire the form of primary transformations but with the parameters $\theta_{[1 2]}$ and $\beta_{[1 2]}$. It is straightforward to demonstrate that, as well as in the case of (\ref{gts})-(\ref{a}), these parameters obey the same group composition law
\begin{eqnarray}
i\theta_{[1 2]}&=& [\theta_1, \theta_2] +[\beta_1, \beta_2] \\
i\beta_{[1 2]}&=& [\theta_1, \beta_2 ]+[\beta_1, \theta_2].
\end{eqnarray}
This is nothing else but the composition law of the $SU(2)_L\times U(1)_R$ group: $i\omega_{[1 2]}=[\omega_1, \omega_2]$, and $\alpha_{[1 2]}=0$.  
Therefore, the change of variables (\ref{noncd}) in the functional integral (\ref{smat3}) does not destroy the gauge invariance of the theory, but rather affects the transformation properties of the vector and axial-vector fields in the asymmetric vacuum. 

This means that, in this case, the non-anomalous part of the effective Lagrangian is also invariant under the action of the $SU(2)_L\times U(1)_R$ gauge group. Thus we have a compelling argument [based on the Chisholm-Kamefuchi-O'Raifeartaigh-Salam theorem] to conclude that both approaches are equivalent in their description of the processes governed by the non-anomalous part of the effective action. In other words, this means that a gauge covariant description hides the dominance picture of spin-1 mesons, but does not change the result.       
 
 There is an obvious, troublesome question. If covariant and non-covariant approaches are equivalent for the non-anomalous interactions, why do they differ in the case of anomalous $a_1\to\gamma\pi\pi$ decay? We expect that the answer is hidden in the different transformations of spin-1 states and requires the study of the general solutions of the Wess-Zumino anomaly equation with spin-1 mesons. Our work contains all necessary elements to resolve the issue and gives a basis for such studies.

\section{Conclusions}
\label{concl}

In this paper we have presented a novel covariant approach to the problem of pseudoscalar - axial-vector diagonalization of the effective meson Lagrangian with electroweak interactions included. We have shown that covariant frameworks do not destroy the standard chiral and gauge transformations of the fields, but add new vertices to the effective Lagrangian. This formulation gives rise to a deviation from the vector and axial-vector meson dominance picture, i.e., the theory together with the typical vector/axial-vector dominance diagrams contains also the diagrams that describe the direct emission of gauge particles by the quark-antiquark pair.  

Special attention has been given to the question of gauge symmetry conservation in the standard non-covariant approach. We have obtained the corresponding transformation laws for vector and axial-vector fields to prove that this approach possesses the property of electroweak gauge invariance after the $p-a_\mu$ diagonalization.  

The non-anomalous part of the effective action has been studied in detail. By invoking the Chisholm-Kamefuchi-O'Raifeartaigh-Salam theorem and gauge symmetry arguments we have argued that the covariant approach is equivalent to the standard non-covariant one, provided that only non-anomalous interactions are considered. Thus, in this specific case, a switch to covariant or non-covariant description does not alter the $S$-matrix. This leads us to conclude that the covariant version hides the spin-1 mesons dominance picture of electroweak interactions, without changing the physical [on mass shell] content of the theory.         

One of the interesting future applications of the obtained theory is the study of its anomaly sector. Our previous investigations have shown that new non-VMD contributions arising in the covariant approach are responsible for the restoration of the gauge symmetry for the $a_1\to\gamma\pi^+\pi^-$ decay amplitude, which is not gauge invariant in the non-covariant approach. It would be also important to clarify the role of the new contributions, which originate in the weak sector and induce deviations of the theory from the axial-vector dominance of the charged hadronic current.

Furthermore, one should mention that the extension of the chiral group to the $SU(3)_L\times SU(3)_R$ case will allow us to apply the idea of gauge invariant $p-a_\mu$ diagonalization to the strange quark physics.

\section*{Acknowledgments}
A.A.O. is grateful for warm hospitality at the IMP of the Chinese Academy of Sciences in Lanzhou. P.M.Z is supported by the National Natural Science Foundation of China (Grant No. 11575254). B.H. acknowledges CFisUC and FCT through the project UID/FIS/04564/2016. We acknowledge networking support by the COST Action CA16201.

\end{document}